\documentclass[12pt]{article}

\usepackage{epsfig}

\newcommand{\be} {\begin{equation}}
\newcommand{\ee} {\end{equation}}
\newcommand{\lan}{\langle}
\newcommand{\ran}{\rangle}

\renewcommand{\d}{{\mbox d}}

\newcommand{\qea}{\mbox{$q_{\scriptscriptstyle {\rm EA}}$}}
\newcommand{\qqea}{\mbox{$q_{\scriptscriptstyle {\rm EA}}^2$}}
\newcommand{\de} {\partial}

\begin{document}

\title{Off-Equilibrium Dynamics at Very Low Temperatures in 3d Spin Glasses}

\author{Enzo Marinari$^{(a)}$, Giorgio Parisi$^{(b)}$,\\
Federico Ricci-Tersenghi$^{(c)}$ and Juan J. Ruiz-Lorenzo$^{(d)}$\\[0.5em]
$^{(a)}$ {\small Dipartimento di Fisica and INFN, Universit\`a di Cagliari}\\
{\small \ \  Cittadella Universitaria, S. P. Montserrato-Sestu Km 0.7
09042 Montserrato (CA) (Italy)}\\[0.3em]
{\small \tt Enzo.Marinari@ca.infn.it}\\[0.5em]
$^{(b)}$ {\small Dipartimento di Fisica and INFN, Universit\`a di Roma}
{\small {\em La Sapienza} }\\
{\small \ \ P. A. Moro 2, 00185 Roma (Italy) }\\[0.3em]
{\small \tt Giorgio.Parisi@roma1.infn.it}\\[0.5em]
$^{(c)}$ {\small Abdus Salam International Center for Theoretical Physics}\\
{\small \ \ Strada Costiera 11, P. O. Box 586, 34100 Trieste (Italy) }\\[0.3em]
{\small \tt Federico.Ricci@ictp.trieste.it }\\[0.5em]
$^{(d)}$ {\small Departamento de F\'{\i}sica Te\'orica I,
Universidad Complutense de Madrid}\\
{\small \ \ Ciudad Universitaria, 28040 Madrid (Spain)}\\[0.3em]
{\small \tt ruiz@lattice.fis.ucm.es}\\[0.5em]
}

\date{October 11, 1999}

\maketitle

\begin{abstract}
We present a high statistic systematic study of the overlap
correlation function well below the critical temperature in the three
dimensional Gaussian spin glass. The off-equilibrium correlation
function has been studied confirming the power law behavior for the
dynamical correlation length. In particular we have computed the
dynamical critical exponent $z$ in a wide range of temperatures, $0.35
\le T \le 0.9$, obtaining a dependence $z(T)=6.2/T$ in a
very good agreement with recent experiments.  Moreover, we report a
study of the violation of the fluctuation-dissipation theorem for very
low temperatures $T=0.5$ and $T=0.35$.  All our numerical results
avoid a droplet model interpretation even when $T$ is  as
low as $T=0.35$.
\end{abstract}  

\thispagestyle{empty}
\newpage

\section{Introduction}

The nature of the low temperature phase of finite dimensional spin
glasses is still a subject 
of controversy \cite{DROPLET,MEPAVI,NS,REPLICAS,MOORE}. 

Recently Bray, Moore, Bokil and Drossel \cite{MOORE,BRAY} have
questioned many of the numerical results obtained with Monte Carlo
methods in the three dimensional Edwards-Anderson (EA) model
\cite{REPLICAS,BOOK,3D_EQUIL,4D_EQUIL,WINDOWS}.

Inspired by the study of the Migdal-Kadanoff approximation (MKA) of
the EA model, they argued that the numerical results, that were obtained
at temperatures $T \ge 3/4\,T_c$, could be strongly affected by finite
size effects and that one should go to sizes larger than the crossover
length $L^*$ in order to see the right (droplet) behavior.  They found
(in the framework of the MKA) that the crossover length is $L^* \simeq
100$ for $T \simeq 0.7\,T_c$ and that it decreases for lower
temperatures: $L^* \le 10$ when $T \le 0.5\,T_c$ \cite{MOORE,BRAY}
(see also the comment \cite{COMM}).

This is maybe the main motivation that pushed us to study the EA model
in the very low temperature region: verify whether the
behavior already found at $T \simeq 0.75\,T_c$ persists at $T \le
0.5\,T_c$.  In fact at these temperatures we can simulate (using
off-equilibrium techniques) a system of size larger than $L^*$
(in this paper, we will present data for sizes $L=24$ and $L=64$).

The numerical data showed in this paper have been measured in the
off-equilibrium dynamical regime.  This way of probing the system
properties, apart from being much more similar to the experimental
procedure, does not present the thermalization problems of a
simulation performed at the equilibrium, that would be insurmountable
obstacles at so low temperatures.  The efficiency of this way of 
measuring has been largely tested in the recent
past \cite{MAPARURI,4DIM,MEANFIELD,H4D,H4D_PM1}.  Moreover the
off-equilibrium dynamics of spin glasses has received in the last
years a great attention both from the experimental \cite{ORBACH,EXPER} and
from the analytical \cite{Leticia} points of view.

Taking the measurements in the off-equilibrium regime we are able to
confront the droplet model (DM) \cite{DROPLET}  
and the Mean Field like theory (MF)\cite{REPLICAS}  on
two grounds: the off-equilibrium regime itself and the equilibrium
one, that can be obtained in the limit of very large
times. We can take this limit quite safely thanks to
the very large time reached in our simulations.

A preliminary analysis based on the data at temperatures $T=0.7$ and
$T=0.35$ was reported in reference \cite{MAPARURI}.  In this paper
we present an extended analysis based on nine different
temperatures obtaining a precise temperature dependence of the
dynamical critical exponent in order to have an accurate confront with
recent experiments.

\section{The model and the numerical method}

We have simulated the Gaussian Ising spin glass on a three-dimensional
cubic lattice of volume $L^3$ with periodic boundary conditions. The
Hamiltonian of the system is
\begin{equation}
{\cal H}=-\sum_{<ij>} \sigma_i J_{ij} \sigma_j\ .
\end{equation}
We denote by $<\!ij\!>$ the sum over nearest neighbor pairs. $J_{ij}$
are Gaussian variables with zero mean and unit variance.

We focus our attention on the study of the point-point overlap
correlation function computed at distance $x$ and time $t$
\begin{equation}
G(x,t) = \frac{1}{L^3} \sum_i\overline{\langle \sigma_{i+x} \tau_{i+x}
\sigma_i \tau_i \rangle_t}\ .
\end{equation}
where $\sigma$ and $\tau$ are two real replicas (systems which evolve 
with the same
disorder) and the index $i$
runs over all the points of the lattice.  As usual we denote by
$\overline{(\cdot \cdot \cdot)}$ the average over the disorder and, in
this context, $\langle (\cdot \cdot \cdot) \rangle_t$ is the average
over the dynamical process until time $t$ (for a given realization of
the disorder).  The two replicas ($\sigma$ and $\tau$) evolve with
different random numbers.

The simulation has been  performed in a similar way to the experimental
procedure: the system is prepared in a high temperature configuration
(actually the initial configurations were chosen at random, i.e.\
$T=\infty$) and suddenly it is quenched below the (estimated) critical
temperature, $T_c=0.95(3)$ \cite{3D_EQUIL}. Immediately we start taking the
measurements, which obviously depend on time.  The equilibrium behavior
is recovered in the large time limit. We have used as dynamical
process the standard Metropolis method.

We have simulated 4 samples (8 systems) of an $L=64$ lattice,
measuring the correlation function at times $t = 100 \cdot 2^k$ (with
$k=0,\ldots,13$) and temperatures $T=0.9$, $0.8$, $0.7$, $0.6$, $0.5$,
$0.4$, and $0.35$.  In addition, we have simulated 4096 samples of an
$L=24$ lattice measuring at times $t=2^k$ (with $k=7,\ldots,19$) and
at three temperatures: $T=0.7$, $0.5$ and $0.35$.

For the study of the fluctuation-dissipation relation we have used
$L=64$ systems and we have simulated them for more than $10^7$ Monte
Carlo steps.  All the simulations have been performed with the help of
the parallel computer APE100 \cite{APE}.

\section{Results on the correlation function}

First, we analyze the correlation functions computed with the
$L=64$ runs.  The study of the numerical data suggest us the
following Ansatz for the time and spatial dependences of the
correlation function \cite{MAPARURI}
\begin{equation}
G(x,t) = \frac{\mbox{\rm const}}{x^\alpha}
\exp \left[ - \left( \frac{x}{\xi(t)} \right)^\delta  \right]\ ,
\label{main}
\end{equation}
where $\xi(t)$ is the dynamical correlation length.  The numerical
data clearly show that the dynamical correlation length depends on the
time following a power law $\xi(t) = B\,t^{1/z}$ where $z$ is the
dynamical critical exponent.  The exponents $\alpha$, $\delta$ and $z$
and the amplitude $B$ could, in principle, depend on the temperature.
However we obtain (see below) that $\alpha$, $\delta$ and $B$ are
almost temperature independent, while $z(T)$ is inversely proportional
to $T$.

\begin{table}
\centering
\begin{tabular}{|c|c|c|c|c|} \hline
$T$  &  $z(T)$    & $\delta$ & $B^{-\delta}$ & $\alpha$\\ \hline \hline
0.9  &  6.85(1.0) & 1.37(11) &    1.02(4)    & 0.60(7) \\ \hline
0.8  &  7.5(1.3)  & 1.37(9)  &    1.06(4)    & 0.49(9) \\ \hline
0.7  &  9.3(0.7)  & 1.50(5)  &    1.02(4)    & 0.53(5) \\ \hline
0.6  & 10.3(1.2)  & 1.38(3)  &    1.09(4)    & 0.49(16)\\ \hline
0.5  & 11.7(1.8)  & 1.43(2)  &    1.04(4)    & 0.59(20)\\ \hline
0.4  & 14.1(2.4)  & 1.45(4)  &    0.99(3)    & 0.60(26)\\ \hline
0.35 & 19.9(3.8)  & 1.41(6)  &    1.03(7)    & 0.29(32)\\ \hline
\end{tabular}
\caption{ Parameters of the point-point correlation function}
\protect\label{table:param}
\end{table}

In Table~\ref{table:param} we report the results of our fits (always
done using the CERN routine MINUIT \cite{MINUIT}).  We remark that,
for a given temperature, we have fitted our numerical data to the
Ansatz of Eq.(\ref{main}) in two steps.  In the first step we fix the
distance in the correlation function and we perform the following
three parameters fit in the variable $t$
\begin{equation}
\log G(x,t)= A(x) - B(x)\,t^{-\delta/z}\ .
\end{equation}
We have found that $\delta/z$ is independent of $x$.  The second
step has been to extract from $A(x)$ and $B(x)$ the exponents $\alpha$ and
$\delta$ and the amplitude $B$ using the formul\ae: $A(x)= \mbox{\rm
const} - \alpha \log x$ and $B(x)= B^{-\delta} x^\delta$. We report
our final values of $z$, $\delta$, $B$ and $\alpha$ in 
Table~\ref{table:param}.

The resulting values for $z(T)$ (see Table~\ref{table:param}) 
can be fitted to a power law
 (using all the temperatures of Table~\ref{table:param}) obtaining
\begin{equation}
z(T)=6.4(6)\,T^{-0.96(20)}\ .
\end{equation}
From the previous fit we can guess a simpler law for the dynamical
critical exponent $z(T)=a/T$, obtaining\footnote{ 
        This law was found by Kisker {\em et al.} for the $\pm 1$
        three dimensional spin glass (See references \cite{Kisker,RIEGER}).
        Moreover this law was guessed for the Gaussian model 
        using numerical data taken at
        temperatures $T=0.7$ and $T=0.35$ in reference \cite{MAPARURI}.}
\begin{equation}
z(T)=\frac{6.2(3)}{T}\ .
\end{equation}
This kind of behavior suggests that the low temperature dynamics in spin
glasses is dominated mainly by activated processes with free energy
barriers diverging logarithmically with the size of the system.  

We can finally write down the dependence of the dynamical
correlation length on the time as well as on the temperature:
\begin{equation}
\xi(t,T) \propto t^{T/6.2(3)} = t^{0.161(8)\ T} = t^{0.153(12)\ T/T_c}\; ,
\label{final}
\end{equation}
where we have assumed that the temperature of the phase transition is
$T_c=0.95(3)$ \cite{3D_EQUIL}.  The agreement of the previous formula with the
experiments is very good.  We recall that in
experiments \cite{ORBACH} it was found the following dependence for
the dynamical correlation length
\begin{equation}
\xi(t,T) \propto t^{0.170 ~T/T_g}\ .
\end{equation}
where $T_g$ is the experimental critical temperature (the authors of
this result do not quote the error in the exponent).

A further check of  Eq.(\ref{main}) would be the
collapse of the data (measured at different times and different
temperatures) when plotting $G(x,t) x^\alpha$ versus $x/t^{1/z(T)}$.
To this purpose we use the data from  4 samples of the $64^3$
runs, together with those measured on 4096 samples of $24^3$
runs.  We remark that, in the $24^3$ runs, the volume is near 19 times
less than the $L=64$ runs but we have computed 1000 times more samples
and so we expect the errors to be smaller.

\begin{figure}
\begin{center}
\leavevmode
\epsfig{file=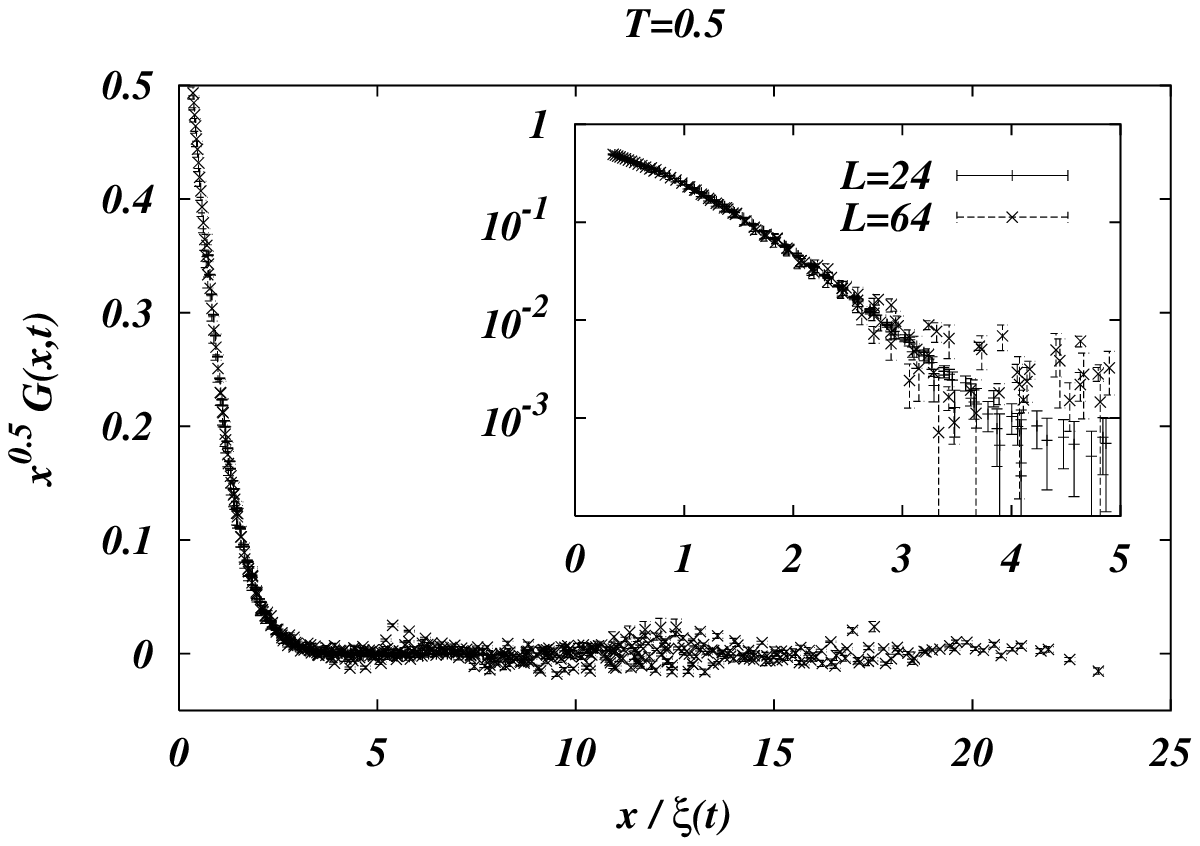,width=0.85\linewidth}
\epsfig{file=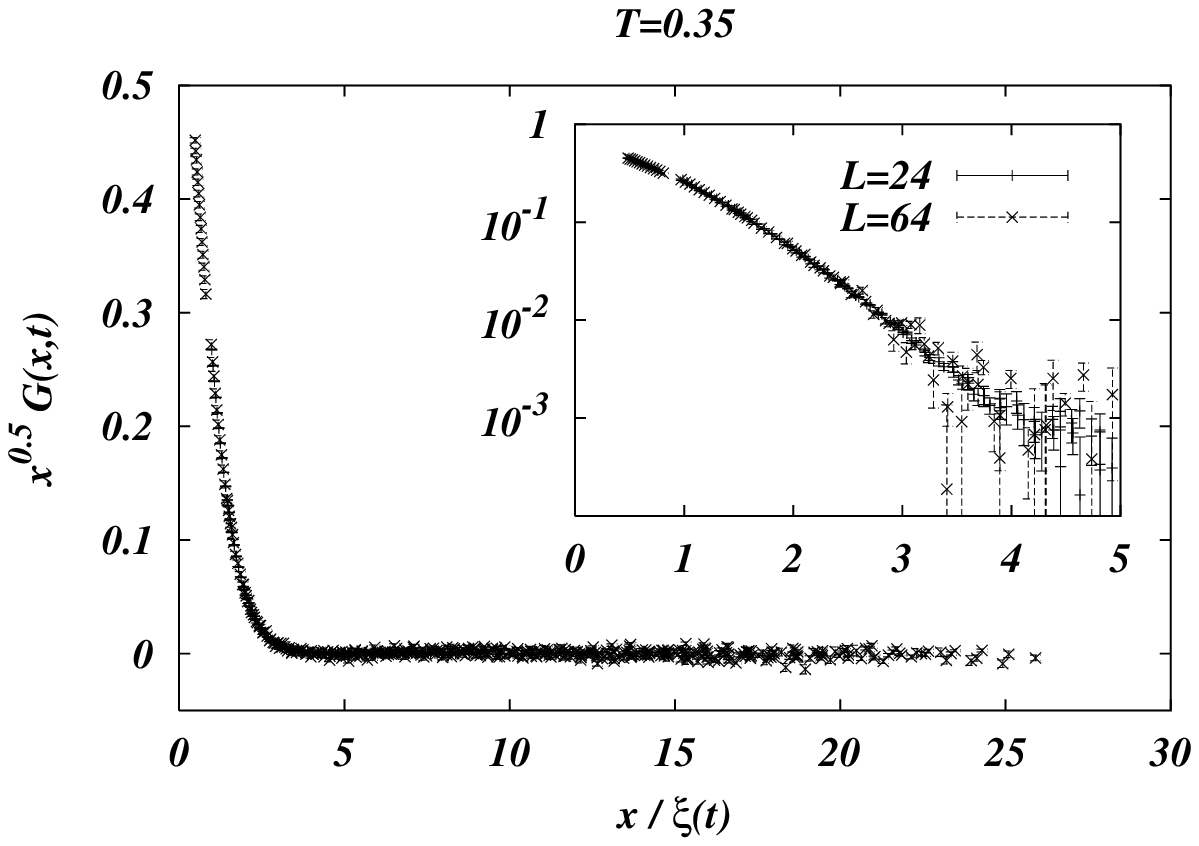,width=0.85\linewidth}
\end{center}
\caption{Scaling plot for the correlation function $G(x,t)$ 
measured at two very low temperatures, $T=0.5$ and $0.35$, and two
lattice sizes, $L=24$ and $64$. It shows that the finite size effects
are negligible and it also gives reliability to our estimate for
$\xi(t)$.}
\label{plot_corr}
\end{figure}

In Fig.~\ref{plot_corr} we plot the correlation function for two low
temperatures ($T=0.5$ and $0.35$) using as variables $x/\xi(t)$ and
$x^\alpha G(x,t)$ (we have taken $\alpha=0.5$, see
Table~\ref{table:param}).  In the plots we use
the data from both runs ($L=24$ and $L=64$) and they superimpose
perfectly.  In the insets we present the same data in a log-linear
scale in order to let the reader evaluate better the collapse.  It is
clear that the scaling is impressive even at the lowest temperature
$T=0.35$. We can also state that the finite size effects are
negligible for the lattice sizes used.

Scaling arguments tell us that the more general scaling function for
the correlation function is (for large $x$ and $t$)
\begin{equation}
G(x,t) \propto x^{-\alpha} {\cal G}\left(\frac{x}{\xi(t)}\right)\ ,
\label{scaling}
\end{equation}
where the scaling function ${\cal G}(y)$ is smooth.  Moreover,
in the scaling regime, ${\cal G}(y)$ should not depend neither on temperature
nor on the lattice size.  Note that in our Ansatz~(\ref{main}) we
have chosen an exponential function for the scaling function: ${\cal
G}(y) \propto \exp(-y^\delta)$, 
and we show that it fits very well the
data. However to check that our estimates of $\alpha$ and $\xi(t)$ are
correct we do not need to know ${\cal G}(y)$.  We can simply
plot $x^\alpha G(x,t)$ versus $x/\xi(t)$ (as it was done in
Fig.~\ref{plot_corr}) and check how well the data collapse.

\begin{figure}
\begin{center}
\leavevmode
\epsfig{file=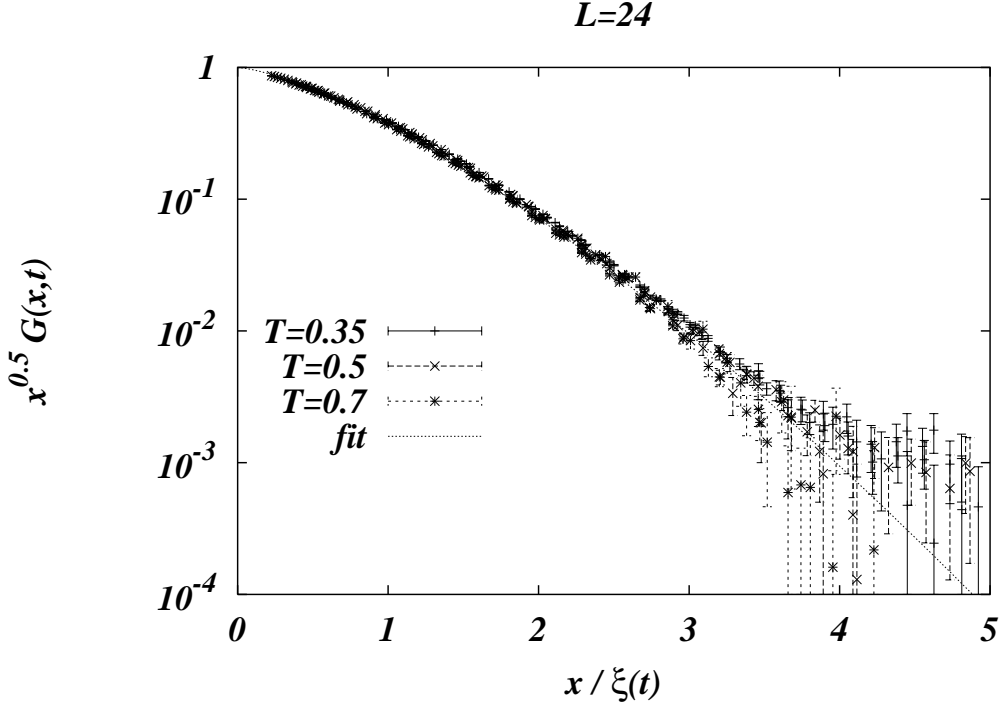,width=0.85\linewidth}
\end{center}
\caption{The scaled data for $G(x,t)$ are well described by a
temperature-independent scaling function, that can be very well
approximated by our fit.}
\label{tutte_T}
\end{figure}

In order to check the temperature independence of ${\cal G}(y)$ we
show in Fig.~\ref{tutte_T} the scaling function for three different
temperatures ($T=0.35,0.5$ and $0.7$), together with the exponential
function $\exp[-y^{1.42(2)}]$ (see Table ~\ref{table:param}) 
obtained through the fitting procedure.
It is clear that the scaling function is really
temperature-independent and it can be very well approximated by the
exponential function as we have chosen in our Ansatz.

Another interesting issue is the extrapolation of the correlation
function to infinite time.  In this limit we can compare again our
numerical results with the predictions of the droplet model and with
that of the RSB theory.  In the former the extrapolated correlation
function tends to the value \qqea\ for large distances, whereas the
RSB prediction is a pure power law going asymptotically to zero 
\cite{DeDominicis}.  
Our Ansatz, which describes
perfectly the numerical data, supports the RSB prediction even for the
lowest temperatures.

Nevertheless, we have tried to fit our data with a functional
dependence compatible with the droplet model, that is,
$G(x,t)=G_\infty(x)\,{\cal G}(x/\xi(t))$, where $G_\infty(x) =
A\,x^{-\alpha} + C$. If $C=0$, then the previous formula is exactly
our Ansatz (and it implies a breaking of the replica symmetry), while
if $C=\qqea$ then it would support a droplet picture.  Fitting the data to
the previous formula, $G_\infty(x)$, we have found that at every
temperature and even at $T=0.35$ (our lowest temperature), the best value for
$C$  is always compatible with zero.  At very low
temperatures, i.e.\ $T=0.35$, the Edwards-Anderson order parameter is
so close to one ($\qea \simeq 1$) than we can safely distinguish
between the two competing theories.  In fact in the droplet like
formula we have that $G_\infty(x)$ is almost 
constant\footnote{
        Actually it slowly decreases from 1 to $\qqea \simeq 1$, but for all
        our purposes it can be considered as a constant.}  
and so we simply
should fit the data into the scaling formula of Eq.(\ref{scaling})
without the factor $x^{-\alpha}$ in order to check the correctness of
the droplet model.

\begin{figure}
\begin{center}
\leavevmode
\epsfig{file=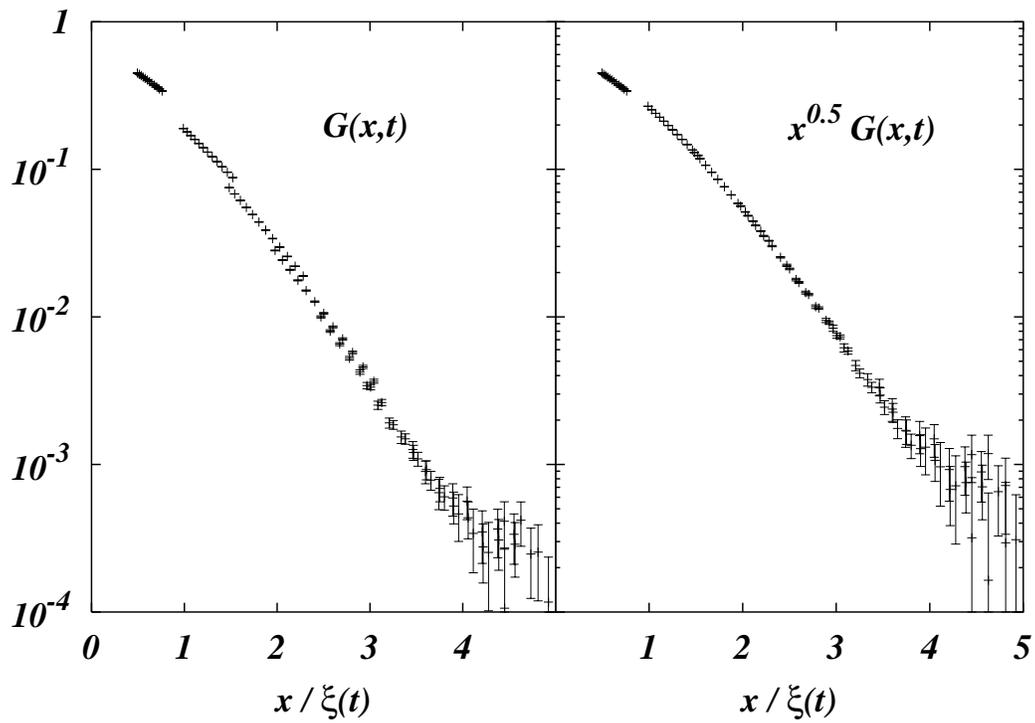,width=0.85\linewidth}
\end{center}
\caption{The comparison of these two scaling plots clearly shows that
the $\sqrt{x}$ factor is essential in order to collapse the
correlation function data. The temperature is $T=0.35$ and the lattice
size $L=24$.}
\label{con_senza}
\end{figure}

In Fig.~\ref{con_senza} we present the data rescaled with the formula
suggested by the droplet model (left plot) and with that implied by
RSB (right plot).  It is clear that the RSB prediction fits much
better the numerical data.  Note that the data error is sufficiently
small to affirm safely that the data in the left plot have no collapse
at all.  A scaling plot like the left one has been recently presented
by Komori et al.\ in \cite{KOMORI} (see also reference
\cite{Comment}).  We believe that the rather poor collapse of their
data (see Fig.\ 5 in \cite{KOMORI}) is due to the fact that they neglect
the factor $x^{-\alpha}$ in the scaling formula.  A much better
collapse would be obtained by plotting $\sqrt{x}\ G(x,t)$ versus
$x/\xi(t)$ (see \cite{Comment}).

\section{Fluctuation dissipation relation at very low temperatures}

Now, we present the results of the analysis based on the
generalization of the fluctuation-dissipation theorem (FDT) in the out of
equilibrium regime \cite{CUKU}.  
In this section we will focus on the scaling
properties of the aging region and the violation of
fluctuation-dissipation at very low temperatures.

A preliminary analysis was done in \cite{FDT} studying the violation of
FDT at temperature, $T=0.7$.  Here we have simulated different lower
temperatures and so, as byproduct, we can study the scaling properties
of the violation of FDT.  An analogous analysis, with many
temperatures, was done in \cite{FDT} but on the four-dimensional EA
model.

For the sake of conciseness, we do not repeat all the formalism and we
address the interested reader to one of the previous publications on
the subject \cite{CUKU,RIEFRA,FDT,FDT_DILU,H4D}.  Here we simply recall the
main formul\ae\ that we use.  As usual we define the integrated
response to a very small external field as
\begin{equation}
\chi(t,t_w) = \lim_{h_0 \to 0} \frac{1}{h_0}
\int_{t_w}^t\,R(t,t')\,h(t')\,\d t'\ \ ,
\end{equation}
where $h(t) = h_0\,\theta(t-t_w)$ and $R(t,t') = \frac1N \sum_i
\frac{\de \lan s_i(t) \ran}{\de h(t')}$. The autocorrelation
function is defined as
\begin{equation}
C(t,t_w) = \frac1N \sum_i \lan s_i(t) s_i(t_w) \ran\ \ .
\end{equation}
Relating these two functions, in the large times limit, via
\begin{equation}
T \chi(t,t_w) = S(C(t,t_w))\ \ ,
\end{equation}
we have that, at the equilibrium, the fluctuation-dissipation theorem
(FDT) holds and $S(C) = 1-C$, while in the aging regime the function
$S(C)$ can be linked to the equilibrium overlap distribution through
$P(q) = -\left . \frac{\de^2 S(C)}{\de C^2} \right|_{C=q}$ \cite{FDT,ANALIT}.

Models that, in the frozen phase, do not show any breaking of the
replica symmetry, have, at the equilibrium level, a static $P(q)
= \delta(q-\qea)$, which dynamically corresponds to the absence of
response in the aging regime.  This means that, plotting $\chi(t,t_w)$
versus $C(t,t_w)$, we obtain a horizontal line in the range $C \le
\qea$ \cite{FDT_DILU} (in the quasi-equilibrium regime, $C \ge \qea$,
and it always holds  $T \chi = 1-C$ independently of the model)

\begin{figure}
\begin{center}
\leavevmode
\epsfig{file=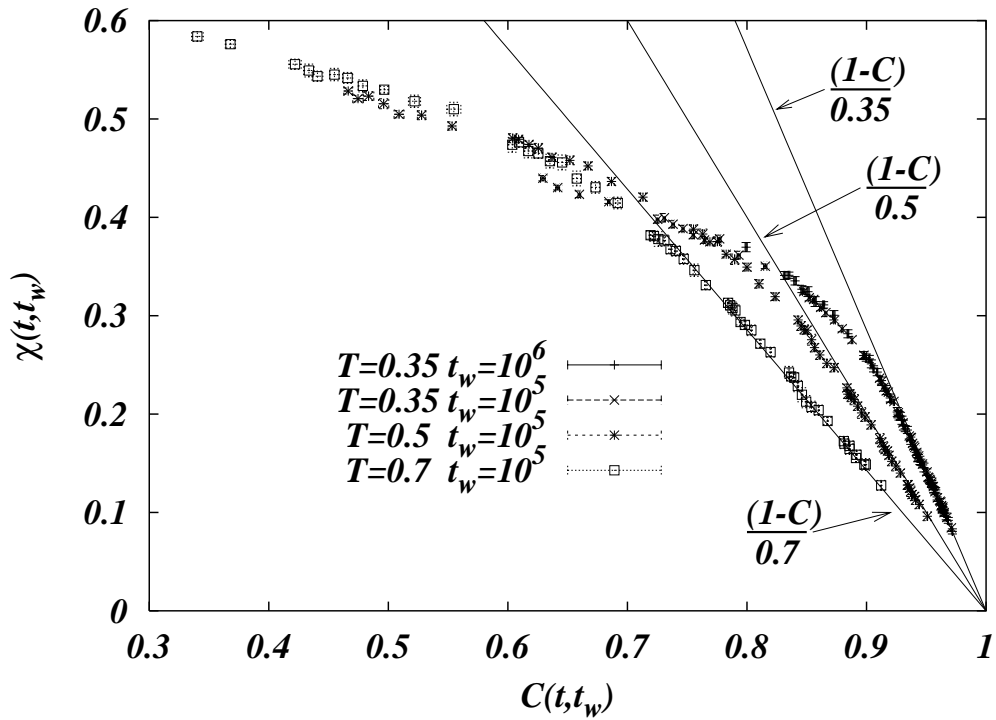,width=0.85\linewidth}
\end{center}
\caption{Response against the autocorrelation function for three
different temperatures and lattice size $L=64$.  
Note that in this figure we plot
$\chi(t,t_w)$ versus $C(t,t_w)$.  The data stay on a single universal
curve when they leave the FDT lines.  This curve is clearly non
horizontal and this hints for a breaking of the replica symmetry in
the very low temperature phase of the EA model.}
\label{FDT}
\end{figure}

In Fig.~\ref{FDT} we show the results for different temperatures in
the usual plot $\chi(t,t_w)$ versus $C(t,t_w)$. Note that in this plot
the FDT line is $\chi = (1-C)/T$ and so it is different for different
temperatures.  It is quite clear that, even for very large times, the
curves are far from been horizontal when they leave the FDT line.
This result gives more evidence in favor of a replica symmetry
breaking in the very low temperature phase of the 3D EA model \cite{FDT}.

We present the data for different temperatures on a single plot in
order to make more evident the fact that the numerical data seem to
stay on the same curve once the system enters into the aging regime, i.e.\
when the points leave the FDT line.  This kind of behavior has been
observed in the four-dimensional EA model \cite{FDT} and it is
reminiscent of the mean-field solution.

Indeed in the SK model, using the Parisi-Toulouse (PAT)
hypothesis \cite{PAT}, it can be shown \cite{FDT} that
\begin{equation}
S(C) = \left\{
\begin{array}{cl}
1-C & {\rm for}\ C \ge \qea(T)\ \ ,\\
T\,\sqrt{1-C} & {\rm for}\ C \le \qea(T)\ \ .
\end{array}
\right.
\end{equation}
The formula can be easily generalized assuming a generic power law
behavior in the aging regime: $S(C) = T\,A (1-C)^B$ (the mean-field
value for the exponent is $B=1/2$).

We use this generalization to fit the data and we obtain very good
results.  The best fit parameters have been estimated from the collapse
of the data reported in Fig.~\ref{FDT} and they are $A \simeq 0.7$ and
$B \simeq 0.41$ (to be compared with the mean-field values $A=1$ and
$B=1/2$, and those obtained for the 4D EA model $A \simeq 0.52$ and $B
\simeq 0.41$  \cite{FDT}).
 
\begin{figure}
\begin{center}
\leavevmode
\epsfig{file=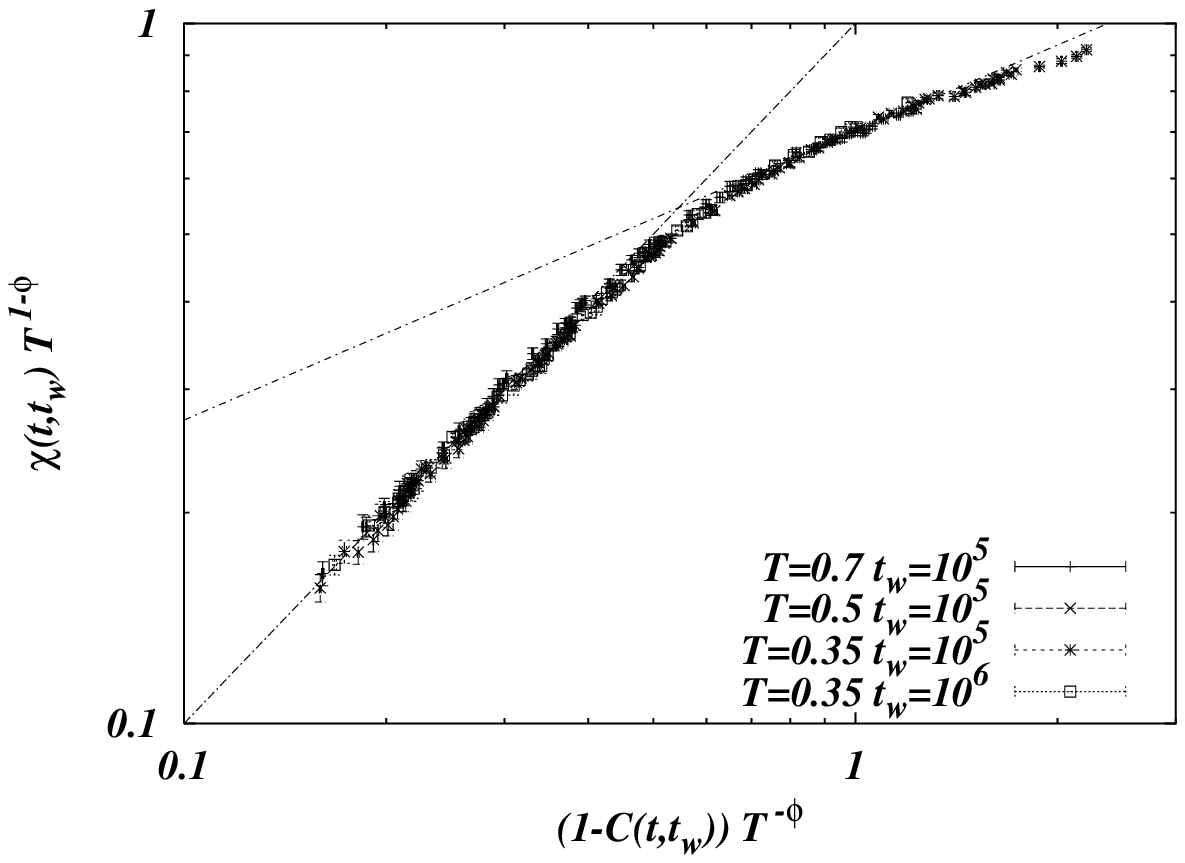,width=0.85\linewidth}
\end{center}
\caption{Scaling plot of the curves shown in the previous figure:
$\chi\,T^{1-\phi}$ versus $(1-C)\,T^{-\phi}$ with $\phi = 1.7$.  In
the FDT regime (left part of the figure) the scaled data stay on the
line $y=x$, while in the off-equilibrium regime (right part of the
figure) they follow the plotted power law $y=A\,x^B$ with $B=0.41$.}
\label{SCALING_FDT}
\end{figure}

In order to show the validity of the fitting formula, we present in
Fig.~\ref{SCALING_FDT} the collapse of the scaled data using the
variables $x=(1-C) T^{-\phi}$ and $y=\chi T^{1-\phi}$, where
$\phi=\frac{1}{1-B}=1.7$.  It is easy to see that, if the previous
scaling holds, the data should stay on two power laws: $y=x$ and $y=A
x^B$ in the quasi-equilibrium and aging regime, respectively.  The two
power laws are reported in Fig.~\ref{SCALING_FDT}.

Even if we may expect a breakdown of the assumed scaling for large
values of the scaling variable $x$ (i.e.\ the scaled data are no
longer described by a power law), we note however that for a quite
large range the collapse is very good and very well approximated by a
power law.  Moreover, we remark that the collapse has been obtained
adjusting only one parameter.

\section{\protect\label{S_CONCLU}Discussion}

We have studied the off equilibrium dynamics of the three dimensional
Gaussian spin glass in the very low temperature phase. In particular
we have studied the scaling properties of the dynamical overlap
correlation functions and the scaling properties of the violation of
the fluctuation-dissipation.

We have tried to fit our correlation functions to the functional form
predicted by the droplet model but the fits were poor. Moreover a
correlation length diverging following a power law with the time
implies, as was noted by Rieger \cite{RIEGER}, barriers diverging not
as $L^\psi$ (as predicted by the droplet model with the lower bound
$\psi \ge \theta \simeq 0.2$) but as $\log L$. This latter results
implies $\psi=0$, hence violating the droplet lower bound. 

It is interesting to note that the experimental data could be fitted
to the droplet formula assuming that $\psi=\theta$ \cite{ORBACH}.
However while both, the results of numerical simulations and the
experiments are in very good agreement with a power law fit for
$\xi(t,T)$, the numerical fit  assuming a droplet formula for
$\xi(t,T)$ \cite{Kisker} disagrees with the experimental 
fit assuming  the same hypothesis \cite{ORBACH}.

As it has been noted above, our final result for the dynamical
correlation length is in a very good  agreement with the
experimental result.

We remark that the same scenario (power law dependence of $\xi(t,T)$
and linear dependence of $1/z$ with temperature) also emerges in four
and six dimensions. In the latter case it was found that $z(T)=4\,T_c/T$
\cite{MEANFIELD} ($z=4$ at the transition is the value predicted by
Mean Field) while in the former one $z(T)=5.5\,T_c/T$ \cite{4DIM}.
Moreover in these two dimensions the overlap correlation function
constrained to zero overlap follows a pure power law as in three
dimensions.

If we send to infinity the time in our Ansatz for the overlap-overlap
correlation function we obtain a pure power decay $G(x) \propto x
^{-\alpha}$ with $\alpha \simeq 0.5$, with a small dependence of
$\alpha$ on the temperature for the whole spin glass phase. We recall
again that the droplet prediction is $G(x) \to \qqea$ in contradiction
with our numerical correlation functions (this fact was already noted
in \cite{MAPARURI}). Instead, the pure power behavior is supported by the
Gaussian approximation using the Mean Field solution 
\cite{DeDominicis,REPLICAS}.

One could argue that the simulated temperatures are not low enough and
the times and sizes not large enough in order to see the ``true''
(droplet) behavior of the EA model.  However, as we stressed in the
Introduction, $L=64$ is large enough for temperatures as low as
$T=0.35$ and $T=0.5$.  Moreover our large times extrapolations are
very safe thanks to the measurements have been taken over 6 time
decades.

We have shown numerical results that contradict the droplet
predictions in a wide range of temperatures ($0.35\le T \le 0.9$).  In
particular we point out that our results (both for correlation
functions and for violation of FDT) at a very low temperature,
$T=0.35$, support a Mean Field picture.

Finally, we remark that using the (PAT) Mean Field scaling relations
for the $P(q)$ \cite{PAT} we have obtained a very good scaling plot of
the violation of fluctuation-dissipation (like in four dimensions
\cite{FDT}). This provides us another strong evidence calling for a
low temperature phase being well described by Mean Field \cite{REPLICAS}.

\section{Acknowledgments}

JJRL is partially supported by CICyT AEN97-1693. JJRL wishes to thank
L.A. F\'ernandez and A. Mu\~noz Sudupe for interesting
suggestions. Moreover we wish to thank A. J. Bray for pointing out an
arithmetic error in equation (\ref{final}).

\end{document}